\documentclass[aps, pra, preprint]{revtex4-2}
\usepackage[T1]{fontenc} 
\usepackage{amsmath} 
\usepackage{bm} 
\usepackage{siunitx} 
\usepackage{booktabs} 
\usepackage{graphicx} 
\usepackage{verbatim} 
\usepackage{listing,lstbayes} 
\usepackage{enumitem} 
\usepackage{url} 

\newcommand{\half}{\frac{1}{2}}

\begin{document}

\title{Comment on ``Physics without determinism: \\ Alternative interpretations of classical physics'', \\ Phys. Rev. A, 100:062107, Dec 2019}
\author{Luca Callegaro, Francesca Pennecchi and Walter Bich}
\affiliation{INRIM - Istituto Nazionale di Ricerca Metrologica \\ Strada delle Cacce, 91 -- 10135 Torino, Italy} 

\begin{abstract}
The paper ``Physics without determinism: Alternative interpretations of classical physics'' [Phys. Rev. A, 100:062107, Dec 2019] defines \emph{finite information quantities} (FIQ). A FIQ expresses the available information about the value of a physical quantity. We show that a change in the measurement unit does not preserve the information carried by a FIQ, and therefore that the definition provided in the paper is not complete.
\end{abstract}

\maketitle
The expression of the state of knowledge about a measurand as a probabilitiy distribution (or some summary of it, such as its mean and standard deviation) is the conventional approach for expressing a measurement result \cite{VIM,GUM,GUM-S1,GUM-S2}. However, it does not intuitively parallel the much more immediate concepts of ``certain'' and ``uncertain digits'' that every experimentalist feels when taking note of a measurement outcome in the lab notebook.

In \cite{DelSanto:2019}, Del Santo and Gisin introduce the concept of \emph{finite information quantities} (FIQ). A FIQ ranging in the interval $[0,1]$ is expressed by the binary number $Q = 0.Q_1 Q_2 Q_3 \ldots$, where the individual bits $Q_k$ are Bernoulli random variables having propensities $q_k$ for the realisation of the case $Q_k=1$. A specific FIQ $Q$ is thus defined by the vector of propensities \mbox{$\bm{q} = [ q_1, q_2, \ldots, q_k, \ldots, q_M, \frac{1}{2}, \frac{1}{2}, \ldots ]$} of its bits $Q_k$; it is assumed that $q_k= \half$ for $k>M$, {\it i.e.}, all bits beyond position $M$ have a $\SI{50}{\percent}$ propensity of being either $0$ or $1$ and therefore carry no information. Only a finite number $M$ of propensities are needed to specify $Q$.

The FIQ concept is very appealing and it is tempting to adopt it to express the value and uncertainty of a quantity as an alternative to probability distributions. However, for the concept of FIQ to become a practical alternative to the current way of representing the state of knowledge about a quantity, it is mandatory that calculations with them be possible and, hopefully, simple.

Consider for example the  expression of the  value of a quantity, traditionally written as  $Q=\{Q\}[U]$, where $\{Q\}$ is the numerical value and $[U]$ is the unit. Changing the unit to $U'= U/L$, $L$ being a constant, implies $Q=\{Q'\}[U']$, with $\{Q'\}=L\{Q\}$. So, even such an elementary tranformation as the change of measurement unit implies the multiplication of a FIQ by a constant.

Indeed, the FIQ definition suggests that it is possible to identify simple, practical calculation rules operating on the finite (and, intuitively, small) number of indeterminate bits and their propensities; rules suitable to be converted in efficient computation algorithms. 

The  arithmetic relevant to a unit change (Appendix A) shows that the transformation $Q'=LQ$ generates bits $Q'_k$ of $Q'$ which are not mutually independent even if the original $Q_k$ bits are independent. Therefore, expressing $Q'$ by providing only the propensities $q'_k$ of its individual bits deletes some of the original information. 

Random variables $Q$ with independent binary digits $Q_k$ have been considered in mathematical literature~\cite{Marsaglia:1971,Pratsevityi:1995,Albeverio:2007}. In general, $Q$ has a `reasonable' probability density function (pdf) only if the $q_k$ satisfy strict conditions, and in that case the pdf is necessarily an exponential~\cite{Marsaglia:1971}; otherwise, it becomes a fractal~\cite{Pratsevityi:1995}, hence difficult to associate with a physical quantity.

In conclusion, it appears that a specification of the state of knowledge about a quantity $Q$ by means of a FIQ should also include information on the dependencies among the $Q_k$, and therefore that, although the FIQ concept might be physically sound and useful, its definition as given in \cite{DelSanto:2019} is not complete, and deserves further development.
\appendix
\section{Minimal FIQ maths}
A FIQ arithmetics can be established by generalizing operations on binary numbers. The sum \mbox{$S=Q+R=0.S_1 S_2 S_3 \ldots$} of two FIQs, \mbox{$Q=0 . Q_1 Q_2 Q_3 \ldots$} and \mbox{$R =0 . R_1 R_2 R_3 \ldots$}, is given by the full adder rule, Tab.~\ref{tab:binaryfulladder}. 
\begin{table}[ht]
\centering
\begin{tabular}{ccc|cc}
\toprule
$Q_k$ & $R_k$ & $C_{k+1}$ & $S_k$ & $C_k$ \\
\colrule
 0 & 0 & 0 & 0 & 0\\ 
 0 & 0 & 1 & 1 & 0\\ 
 0 & 1 & 0 & 1 & 0\\ 
 0 & 1 & 1 & 0 & 1\\ 
 1 & 0 & 0 & 1 & 0\\ 
 1 & 0 & 1 & 0 & 1\\ 
 1 & 1 & 0 & 0 & 1\\ 
 1 & 1 & 1 & 1 & 1\\ 
\botrule 
\end{tabular}
\caption{Binary full adder truth table. $C_k$ is the carry bit. \label{tab:binaryfulladder}}
\end{table}

If $\bm{q}$ is the vector of propensities associated with $Q$, and $\bm{r}$ with $R$, then under the assumption of independence of $q_k$ and $r_k$, the propensity $s_k$ of each sum bit $S_k$ can be written as the sum of the four propensities of the $S_k=1$ cases in Tab.~\ref{tab:binaryfulladder}:
\begin{align}
s_k  = & (1-q_k)(1-r_k)c_{k+1} + (1-q_k)r_k(1-c_{k+1}) \nonumber \\ &  + q_k(1-r_k)(1-c_{k+1}) + q_k r_k c_{k+1} \nonumber \\
= &  q_k + r_k + c_{k+1} \nonumber \\ &  - 2(q_k r_k + q_k c_{k+1} + r_k c_{k+1}) + 4 q_k r_k c_{k+1}
\end{align}
and similarly the propensity $c_k$ of the carry bit $C_k$ is
\begin{equation}
c_k = q_k r_k + q_k c_{k+1} + r_k c_{k+1} - 2 q_k r_k c_{k+1}
\end{equation}
For example for the case $c_{k+1} = \half$, we have $s_k=\half$ and $c_k=\half (q_k + r_k)$: the information provided by $q_k$ and $r_k$ is transferred, through the carry bit $C_k$, to bit $S_{k-1}$.

Multiplication by a deterministic constant $L$ can be performed by repeated shifting and addition. Table \ref{tab:by11} gives a simple example.
\begin{table}[ht]
\centering
\begin{tabular}{c c|c|c|c|c}
\toprule
            & $0.$ & $0$ & $0$ & $Q_3$ & \ldots \\
 $\times$   &      &     &     1 & 1 &  \\
\colrule
            & $0.$ & $0$ & $0$ & $Q_3$ & \ldots \\
$+$         & $0.$ & $0$ & $Q_3$ & $Q_4$ &\ldots \\
\botrule 
$=$         & $0.$ & $P_1$ & $P_2$ & $P_3$ &\ldots \\
\end{tabular}
\caption{Multiplication table, $P = L Q$ where $Q=0.0 Q_2 Q_3 \ldots$ and $L = (11)_2 = (3)_{10}$. \label{tab:by11}}
\end{table}
If $P = L Q$, where $\bm{q} = [0, 0, q_3, \half \ldots]$ and $L=(11)_2 = (3)_{10}$, then
\begin{align}
    p_1 & = \frac{1}{2} q_3^2 + \frac{1}{4} q_3, \nonumber \\
    p_2 & = q_3 - q_3^2 + \frac{1}{4}, \nonumber \\
    p_3 & = \frac{1}{2}, \quad \ldots 
\end{align}
The propensity of occurrence of specific digit couples can also be computed. For example, denoting as $p_{12}$ the propensity of the event $\{ P_1=1, P_2=1 \}$ we have $p_{12} = 0$ (to have $P_1=1$, it should occur that $Q_3=1$ and $C_3=1$ at the same time, hence  $C_2=1$. However, the case $\{Q_3=1, C_3=1\}$ always generates $P_2=0$, so $\{ P_1=1, P_2=1 \}$ is never possible). Since \mbox{$p_{12} = 0 \neq p_1 p_2$}, bits $P_1$ and $P_2$ are not independent.
%

\end{document}